\documentclass[journal,12pt,onecolumn,draftclsnofoot,]{IEEEtran} 

\usepackage{graphicx}
\usepackage{amsmath}
\usepackage{algorithm2e}
\usepackage[english]{babel}
\usepackage[utf8]{inputenc}
\usepackage{comment,xcolor,color}  
\usepackage{array}
\usepackage{multirow}
\usepackage{color,xcolor}
\usepackage{xcolor,wrapfig}
\usepackage{amsmath}
\usepackage{amssymb}
\usepackage{tabularx}
\usepackage{epsfig}
\usepackage{graphicx}
\usepackage[T1]{fontenc}
\usepackage{epstopdf}
\usepackage{subcaption}
\begin{document}

\def\BibTeX{{\rm B\kern-.05em{\sc i\kern-.025em b}\kern-.08em
    T\kern-.1667em\lower.7ex\hbox{E}\kern-.125emX}}

\def \colone{1}
\def \coltwo{2}
\def \ncols{2}

\ifx \ncols \coltwo
\renewcommand{\baselinestretch}{.95}
\fi
    
\newtheorem{theorem}{Theorem}
\newtheorem{lemma}{Lemma}
\newtheorem{proposition}{Proposition}
\newtheorem{corollary}{Corollary}
\newtheorem{definition}{Definition}
\newtheorem{example}{Example}
\title{A Unified Framework for Joint Mobility Prediction and Object Profiling of Drones in UAV Networks}

\author{\IEEEauthorblockN{Han Peng\IEEEauthorrefmark{1}, Abolfazl Razi\IEEEauthorrefmark{1}, Fatemeh Afghah\IEEEauthorrefmark{1}, Jonathan Ashdown\IEEEauthorrefmark{2}}\\
\IEEEauthorblockA{ \IEEEauthorrefmark{1} 
School of Informatics, Computing and Cyber Systems, Northern Arizona University, Flagstaff, AZ, \\ \IEEEauthorrefmark{2} Air Force Research Laboratory, Rome, NY}}

\maketitle

\begin{abstract}
In recent years, using a network of autonomous and cooperative unmanned aerial vehicles (UAVs) without command and communication from the ground station has become more imperative, in particular in search-and-rescue operations, disaster management, and other applications where human intervention is limited. In such scenarios, UAVs can make more efficient decisions if they acquire more information about the mobility, sensing and actuation capabilities of their neighbor nodes. 
In this paper, we develop an unsupervised online learning algorithm for joint mobility prediction and object profiling of UAVs to facilitate control and communication protocols. The proposed method not only predicts the future locations of the surrounding flying objects, but also classifies them into different groups with similar levels of maneuverability (e.g. rotatory, and fixed-wing UAVs) without prior knowledge about these classes. This method is flexible in admitting new object types with unknown mobility profiles, thereby applicable to emerging flying Ad-hoc networks with heterogeneous nodes.

\end{abstract}


\IEEEpeerreviewmaketitle

\ifx \ncols \coltwo
\vspace{-0.2 in} 
\fi

\section{Introduction}

\textcolor{black}{Recently, the use of unmanned aerial vehicles (UAVs) has increased rapidly for many applications including transportation \cite{thiels2015use}, traffic control \cite{kanistras2015survey}, remote sensing \cite{everaerts2008use}, wild-life monitoring \cite{xu2016internet}, smart agriculture \cite{freeman2015agricultural}, surveillance \cite{wall2011surveillance}, broadband satellite communication enhancement \cite{joo2018low} and reconnaissance and border patrolling \cite{girard2004border}.}  
According to the federal aviation administration (FAA), more than 1 million drones are registered with the federal government in 2018 \cite{faa_2016}. 
In some applications, completing intricate tasks is not feasible with a single UAV due to drones' limited flight time, payload and communication range \cite{gupta2016survey}. 
In these situation, often deploying a network of drones is unavoidable. Also in a more time-sensitive applications such as search and rescue, using networked UAVs significantly raises the chance of mission success \cite{quaritsch2010networked}.

An important property of UAV networks is their extremely dynamic network topologies due to freely flying drones \cite{gupta2016survey}.
This becomes even a more challenging issue for the futuristic autonomous UAV networks \cite{rosati2016dynamic,kaleem2018amateur}. 
The dynamic topology of UAV networks, especially when they are composed of  heterogeneous nodes with different levels of maneuverability, reliability, sensing, actuation and communication capabilities calls for a new generation of control, communication, and navigation mechanisms that meet the requirements of these networks \cite{Afghah_ACC18,Razi_Asilomar17}. \textcolor{black}{In particular, it paves the road for providing connectivity and seamless communication through proactive and predictive routing algorithms in order to improve the network operational performance \cite{7820569,6182876,6182872,zhang2014hdre}.}

\begin{figure}[t]
\centering
\ifx \ncols \colone
\includegraphics[width=0.55\columnwidth]{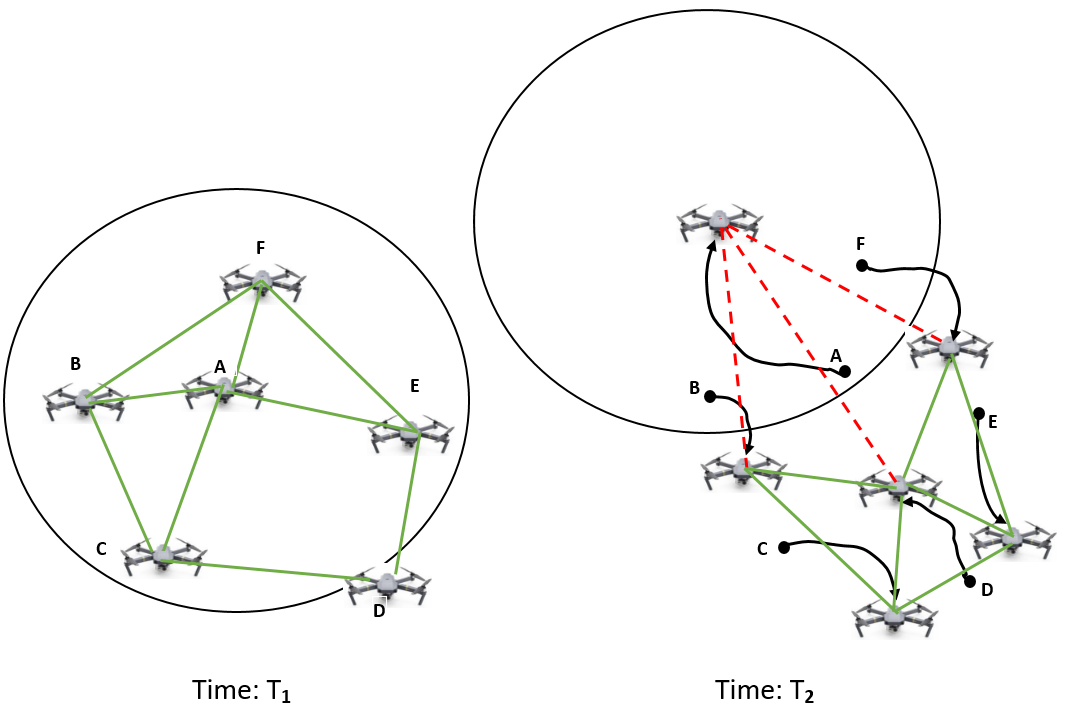}
\else
\includegraphics[width=.75\columnwidth]{Fig1Connectivity.png}
\fi
\caption{\footnotesize Illustration of UAV networks at two time points $T_1$ and $T_2$. The communication range of UAV A is shown by a circle. At time $t=T_2$, A flies away of its neighbors accessible ranges and loses network connectivity. Predicted network topology can be used to prolong network connectivity by selecting routes that are less likely for upcoming failures (e.g. by excluding A).}
\label{fig:coverage}
\ifx \ncols \coltwo
\vspace{-0.3 in} 
\else
\vspace{-0.2 in}
\fi
\end{figure}

Characterizing network topology changes can significantly improve the operational performance of these networks in terms of control and communications, as demonstrated with the following scenarios. For instance, when the network is composed of nodes with limited communication ranges, predicting network topology and the future positions of nodes can be used to enhance network connectivity by excluding the links that are more prone to failure in predictive routing algorithms, as depicted in Fig. \ref{fig:coverage}.
This approach is in a clear contrast with conventional link selection algorithms, where end-to-end routes are set up solely based on the current network topology and link failures are dealt with only after their occurrence. Therefore, the network can suffer from frequent link interruptions and re-establishments \cite{linkselection,khaledi2018greedy}.  
Recently, efforts have been made to develop algorithms for \textit{predictive communication}, with the main idea of making decisions at different layers of communication protocols by taking into account the anticipated future network topology \cite{ArnauWiSEE}.
This new approach of communication requires network topology prediction through member nodes' motion trajectory prediction.

Another example is search-and-rescue operation by autonomous UAV nodes. Predicting the local sub-network topology changes can help each individual UAV to take more efficient decisions. For instance, an autonomous UAV in a search operation may decide to cover areas that  are not already covered and are less likely to be covered by other UAVs based on their predicted motion trajectories. Therefore prediction of node mobility patterns facilitates a more efficient and timely service by autonomous UAVs as depicted in Fig. \ref{fig:disaster}.

\ifx \ncols \coltwo
\vspace{-0.1 in}
\fi

\section{Related Work}
Network topology prediction can be realized by predicting motion trajectories of individual nodes. Here, we assume that UAVs are autonomous with no prior path planning. Also, the UAVs are not allowed to convene and share their current locations and future motion trajectories with one another due to security considerations or limited communication resources. Therefore, each UAV intends to predict the motion trajectories of its neighbor nodes based on its own observation.

Several mobility prediction methods have been proposed in the past decade. These methods can be divided into two main categories of \textit{data-driven} and \textit{model-based} methods. \textit{Data-driven} methods require large datasets to extract the most frequent patterns \cite{Data_driven}. These methods indirectly capture the influence of natural and man-made textures, users' behavioral habits, and spatial and temporal variations on the nodes' mobility \cite{TAPASCologne1}. 
On the other hand, \textit{model-based} mobility models try to predict the motion trajectory of an object based on its motion history and typically rely on the smoothness of motion trajectories~\cite{Model-based}. These methods include piece-wise segment methods \cite{choi2006learning}, hidden Markov models (HMM)~\cite{bennewitz2005learning}, levy flight process \cite{gonzalez2008understanding}, Bayesian methods \cite{aoude2011mobile}, manifold learning \cite{lee2012identification}, and mixture Gaussian models \cite{aoude2011mobile}. These methods are typically customized for specific object types such as human \cite{human}, self-propellers \cite{self-propellers}, and articulated rovers \cite{rovers}.

\begin{figure}[t]
\centering
\ifx \ncols \colone
\includegraphics[width=0.75\columnwidth]{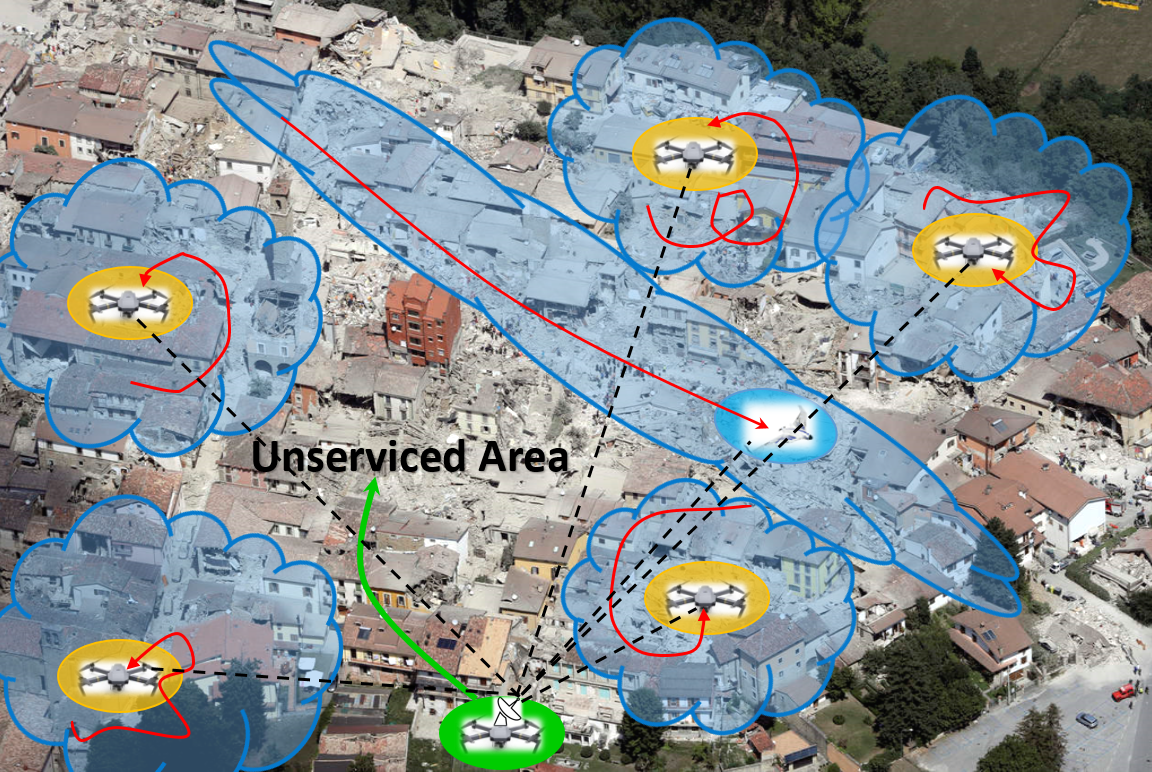} 
\else
\includegraphics[width=.85\columnwidth]{Fig2Disaster.png} 
\fi
\caption{\footnotesize A network of autonomous UAVs perform a search and rescue operations after a natural disaster. A new UAV shown by green color joins the mission. This UAV processes other UAVs motion trajectories to identify and cover the regions that are less likely to be covered by other nodes.}
\label{fig:disaster}
\vspace{-0.2 in}
\end{figure}

The main objective of this work is to develop a unified framework suitable to predict motion trajectories of mobile entities of different types. The core of our method is based on Kalman filtering with intermittent observation \cite{jia2013target}.   
However, we use the object's type-specific motion properties to improve the prediction accuracy through deploying a novel generative model for the system input. 
Thereby, the utilized state transition model provides model flexibility and generality, while class-specific input enables further prediction accuracy. 

The second and more important feature of the proposed method is motion-based object profiling. We note that the predicated node locations in a fully autonomous network are only valid for a near future (a few seconds). Therefore, we need more general and perpetual information about the nodes’ mobility in the majority of applications. 
Here, we visit the network topology prediction problem from a different viewpoint. Note the fact that flying Adhoc networks (FANET) typically include a wide range of object types including ground vehicles, fixed-wing drones, multi-rotor drones, helicopters, and piloted aircrafts, where each type has a different mobility profile. Here, we intend to exploit the main properties of their mobility and classify the objects based on these properties. This approach is inspired by a human perception in recognizing different object types by observing their motion patterns. 
This approach can be used to gain long-term information about the future network topology. For instance, it can be used to predict the coverage area of a UAV in a search and rescue operation as shown in Fig. \ref{fig:disaster}.

To summarize, the objective of this project is to develop a unified framework which jointly performs the two tasks of predicting the near future locations of target nodes as well as classifying them into disjoint types with different maneuverability levels based on their motion profiles. We call this algorithm as joint mobility prediction and profiling (JMPP). 
The proposed system is equipped with a self-tuning module which learns new mobility classes over time without any prior information. This feature provides the flexibility of accepting new object types with new mobility profiles.

It is noteworthy that there are a few recent works focused on clustering motion trajectories using different methods including distance-based clustering \cite{sharma2016trajectory}, 
waypoint clustering \cite{gariel2011trajectory}, tree-based methods \cite{yuan2012efficient}, grid-based methods \cite{mao2017adaptive}, and kernel methods \cite{xu2015unsupervised}. Some of these methods focus specifically on airspace monitoring \cite{gariel2011trajectory}. However, a majority of these methods aim at exploiting the most frequently used geographical paths by mobile objects based on distance metrics, rather than the finding different motion classes. See \cite{yuan2017review} for a more complete review of distance-based methods. Recently, more elegant methods are developed to cluster trajectories based on their shape parameters and not only based on their Euclidean distances. These methods include mixture of multivariate Von Mises distributions \cite{mcfadyen2016aircraft}, sparse nonnegative matrix factorization \cite{pires2017shape}, and circular statistics \cite{mcfadyen2016aircraft}. However, these methods try to find explicit similarities between motion patterns of similar objects which might be absent for most cases. In this work, we approach this problem from a deep learning perspective through exploiting the underlying parameters governing the motion dynamics of an object and use it for object profiling. 

The closest work we have found in the literature is \cite{mcfadyen2016aircraft}, where aircraft motion trajectories are used to classify them into typical manned and expected unmanned aircrafts. They used a trajectory re-sampling technique followed by a mixture of Von Mises distributions to model the trajectories, which are finally clustered using k-medoids algorithm. This method is offline and requires a relatively large dataset of labeled  trajectories and is not capable of performing mobility prediction and online clustering. Further, it is limited to two-classes and is not flexible enough to admit new object types. Our proposed method solves these two important issues, following the recent trend of utilizing advanced machine learning methods in optimizing wireless networking \cite{valehi2017maximizing,valehi2018online,jiang2017machine,alsheikh2014machine}. 



\ifx \ncols \coltwo
\vspace{-0.1 in}
\fi
\section{Universal Mobility Model}   \label{sec:model} 
\ifx \ncols \coltwo
\vspace{-0.05 in}
\fi
In this paper, we view the nodes' mobility from an observer's perspective, which can be any of the network nodes that is monitoring its surrounding partners. 
The kinematic equations of the targets in \textcolor{black}{3D-space} are expressed in terms of state transition equations with a noise term to capture motion turbulence as follows: 
\ifx \ncols \coltwo

\fi
\begin {align} \label{eq:model}
\begin{cases}
\mathbf{x}(k+1)=A\mathbf{x}(k)+F\mathbf a(k)+\mathbf w(k),\\
\mathbf{z}(k) = H\mathbf x(k) +\mathbf \zeta (k),
\end{cases}
\end{align}
\textcolor{black}{where $\mathbf{x}(k) = [x(k) ~ y(k) ~ z(k) ~ v_x(k) ~  v_y(k) ~v_z(k)]^T$ is the state vector (representing the location and velocity of the object at time $k$) and $\mathbf{z}(k) = [z_x(k) ~ z_y(k)~ z_z(k)]^T$ is the observation vector obtained using an arbitrary tracking system. The matrices }
\ifx \ncols \coltwo
{\color{black} \begin{align} 
\nonumber
&A = \begin{bmatrix}  
\mathbf{I}_{3 \times 3}  & dt\mathbf{I}_{3 \times 3} \\
\mathbf{0}_{3 \times 3} & \mathbf{I}_{3 \times 3}
\end{bmatrix},
\\
&F= \begin{bmatrix}  
\mathbf{0}_{3 \times 3} & \mathbf{I}_{3 \times 3}\end{bmatrix}^T, H=\begin{bmatrix} \mathbf{I}_{3 \times 3} & \mathbf{0}_{3 \times 3} \end{bmatrix}^T
\end{align}}
\else
{\color{black} \begin{align} 
&A = \begin{bmatrix}  
\mathbf{I}_{3 \times 3}  & dt\mathbf{I}_{3 \times 3} \\
\mathbf{0}_{3 \times 3} & \mathbf{I}_{3 \times 3}
\end{bmatrix},
&F= \begin{bmatrix}  
\mathbf{0}_{3 \times 3} & \mathbf{I}_{3 \times 3}\end{bmatrix}^T, H=\begin{bmatrix} \mathbf{I}_{3 \times 3} & \mathbf{0}_{3 \times 3} \end{bmatrix}^T
\end{align}}
\fi
define the system, where $dt$ is the time step of the discretized system. Also, $\mathbf{w}(k)\sim \mathcal{N}(\mathbf{0}, \mathbf{R})$ and $\mathbf{\zeta}(k)\sim \mathcal{N}(\mathbf{0}, \mathbf{Q})$ are used to model the system and measurement noise terms.
The key role player, here, is the input vector \textcolor{black}{$\mathbf{a}(k)=[a_x(k) ~a_y(k) ~a_z(k)]^T$}, which represents the acceleration (or equivalently the mechanical forces that drive the whole system dynamics). Therefore, it can be used to define an object's \textit{motion profile}. 

Inspired by the fact that the kinematics of most man-made objects are controlled by acceleration/braking and steering mechanisms, it is desirable to divide the velocity vectors into speed and direction terms as follows:
\ifx \ncols \coltwo
{\color{black}\begin{align} \label{eq:v-w}
\nonumber
&v_{xy}(k)=\sqrt{v_{x}(k)^2+v_{y}(k)^2},v(k)=\sqrt{v_{xy}(k)^2+v_{z}(k)^2},
\\
&w_{\theta}(k)=\theta(k+1)-\theta(k) = \text{tan}^{-1}\frac{v_{y}(k+1)}{v_{x}(k+1)}-\text{tan}^{-1}\frac{v_{y}(k)}{v_{x}(k)},
\end{align}}
\else
{\color{black}\begin{align} \label{eq:v-w}
&v_{xy}(k)=\sqrt{v_{x}(k)^2+v_{y}(k)^2},v(k)=\sqrt{v_{xy}(k)^2+v_{z}(k)^2},
&w_{\theta}(k)=\theta(k+1)-\theta(k) = \text{tan}^{-1}\frac{v_{y}(k+1)}{v_{x}(k+1)}-\text{tan}^{-1}\frac{v_{y}(k)}{v_{x}(k)}
\end{align}}
\fi
\textcolor{black}{where $\theta(k)$ is the direction of the motion trajectory, and $w_{\theta}(k)$ is the angular velocity; both in xy-plane}. Similarly, we can find the linear acceleration in direct path $a(k)$ and the angular acceleration $a_{\theta}(k)$. 
\textcolor{black}{The motion in the 3rd dimension (z-axis) can be considered independent of the motion in xy-plane \cite{xie2013comprehensive}. When the drones hovering in a fixed altitude, we have $v_z(k)=0, a_z(k)=0 \Rightarrow v(k)=v_{xy}(k),~a(k)=a_{xy}(k),$ and the simplified 2D equations can be used \cite{fotouhi2017dronecells,wang2010novel,bouachir2014mobility}.} 
Noting the fact that $a(k)$ and $a_{\theta}(k)$ are time series typically composed of sporadic positive and negative pulses with random amplitudes, we define the following generative model for the system input:
{\color{black}
\begin {align}  \label{eq:acceleration}
\nonumber
&a_i^{xy}(k)\sim(1-\lambda^{xy}_i)\delta(0) +  \lambda_i^{xy}\mathcal{N}(\mu_i^{xy},\sigma_i^{2xy}),\\
\nonumber
&a_i^{z}(k)\sim(1-\lambda^{z}_i)\delta(0) +  \lambda_i\mathcal{N}(\mu_i^{z},\sigma_i^{2z}),\\
&a_i^{\theta}(k)\sim(1-\lambda_i^\theta)\delta(0)+\lambda_i^\theta\mathcal{N}(\mu_i^\theta, \sigma_i^{2\theta}), 
\end{align}}
\textcolor{black}{where $\lambda_i^{xy}$, $\lambda_i^{z}$ and  $\lambda_i^{\theta}$ are Bernoulli distributed random variables (RVs) representing the probability of change, respectively in the velocity in $xy$ plan, the velocity in $z$ direction and the angular velocity in $xy$-plane. Likewise, the amount of change in the speed in $xy$-plane and $z$-axis and the angular velocity in $xy$-plane are modeled with three Gaussian distributions with means $\mu_i^{xy}$, $\mu_i^{z}$ and $\mu_i^{\theta}$, and variances $\sigma_i^{2xy}$, $\sigma_i^{2z}$ and ${\sigma_i^{2\theta}}$}. The subscript $i$ is the object identification (i.d.). This model can be viewed as a special case of \textit{spike and slab} distribution, where the variance of one component approaches zero. Note that this model yields an exponential distribution with a desired memory-less property for the silent intervals between consecutive pulses. The model parameters, denoted by 
\textcolor{black}{$\Theta_i=\{(\lambda^{xy}_i,\mu_i^{xy},\sigma_i^{2xy}),(\lambda^{z}_i,\mu_i^{z},\sigma_i^{2z}),(\lambda_i^{\theta},\mu_i^{\theta},{\sigma_i^{\theta}}^2)\}$} fully determine the statistical properties of the motion dynamics in (\ref{eq:model}). These parameters differ from one object to another, but share similarities among objects within a class. 

\begin{figure}[t]
\centering
\ifx \ncols \colone
\includegraphics[width=0.65\columnwidth]{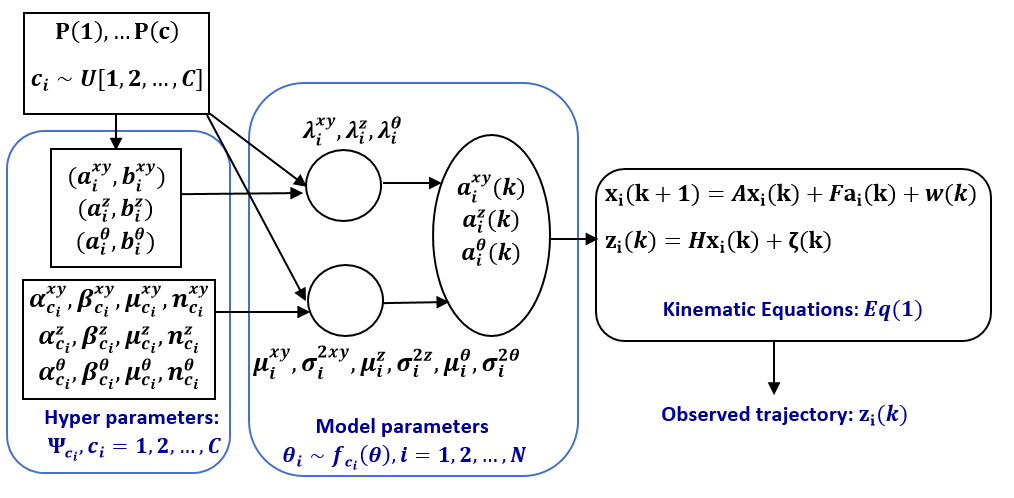}
\else
\includegraphics[width=0.85\columnwidth]{Fig3GraphicalModel3D.png}
\fi
\caption{The proposed mobility model, which includes a universal kinematics model with a probabilistic input represented by a hierarchical graphical model.}
\label{fig:model}
\vspace{-0.2 in}
\end{figure}

To embrace this fact, we model $\Theta_i$ as a random vector whose elements are controlled by a set of hyper-parameters 
\textcolor{black}{$\Psi_{c_i}=\{(a^{xy}_{c_i},b^{xy}_{c_i},\alpha^{xy}_{c_i},\beta^{xy}_{c_i},\mu^{xy}_{c_i},n^{xy}_{c_i})$, $(a^{z}_{c_i},b^{z}_{c_i},\alpha^{z}_{c_i},\beta^{z}_{c_i},\mu^{z}_{c_i},n^{z}_{c_i}),(a_{c_i}^{\theta},b_{c_i}^{\theta},\alpha_{c_i}^{\theta},\beta_{c_i}^{\theta},\mu_{c_i}^{\theta},n_{c_i}^{\theta})\}$}
 shared among objects of class $c_i \in \{1,2,\dots,C\}$. We omit subscript $c_i$ for notation convenience, when it is clear from the text. This approach captures within-class similarities, while providing sufficient flexibility for per-object variability. Fig. \ref{fig:model} provides graphical representation of this generative model for driving forces.

An appropriate choice for model parameters are conjugate distributions, which provide the convenience of closed-from posterior distributions, when applying Bayes' rule. More specifically, the posterior distribution of the model parameters, after observing the acceleration vector \textcolor{black}{$P(\Theta_i|\{a^{xy}_i(k),a^z_i(k),a_i^{\theta}(k)\}_{k=0,1,2\dots})$}, belongs to the same family of prior distribution $P(\Theta_i)$. The Gaussian family is conjugate to itself (or self-conjugate) with respect to a Gaussian likelihood function; thereby its mean can be represented with a Gaussian distribution. The variance also is represented with an Inverse Gamma distribution. The Bernoulli distribution has Beta distribution as its conjugate~\cite{BiShop-Book}. Therefore, we choose the following prior distributions for the model parameters:
{\color{black}
\begin {align} \label{eq:prior}
&\begin{cases}
\lambda_i^\zeta \sim \text{Beta}(a_{c_i}^\zeta,b_{c_i}^\zeta),
\\
\tau_i^\zeta=1/\sigma_i^{2\zeta} \sim \text{Gamma}(\alpha_{c_i}^\zeta, \beta_{c_i}^\zeta),\\
\mu_i^\zeta  \sim \text{N} (\mu_{c_i}^\zeta,  \sigma_{c_i}^{2\zeta}/n_{c_i}^\zeta) & \text{for } \zeta=xy,z, \theta,
\end{cases}
\end{align}}
where $\tau_i$ is the precision and $n_{c_i}$ is an arbitrary shrinkage parameter. Here, we assume that each object $i$ belongs to one class $c_i\in\{1,2,\dots,C\}$. Each class $c$ includes nodes with shared hyper parameter $\Psi_c$. The details of the proposed mobility modeling with probabilistic hierarchical input are presented in Fig. \ref{fig:model}.

Finding the most likely object class based on its observed motion trajectory involves the following steps as depicted in Fig. \ref{fig:method}. Firstly, we estimate the object's current and upcoming locations by solving the state transition equations in (\ref{eq:model}). \textcolor{black}{This stage also provides an estimate of the system acceleration process $\{\hat{\mathbf{a}}_i(k)\}=\{ \big(\hat{a}_i(k),\hat{a}_i^\theta(k)\big)\}_{k=1,2,\dots}$ for a 2D motion and equivalently $\{\hat{\mathbf{a}}_i(k)\}=\{ \big(\hat{a}_i^{xy}(k),\hat{a}_i^{z}(k),\hat{a}_i^\theta(k)\big)\}_{k=1,2,\dots}$ for a 3D motion. Hereafter, we assume $a^z(k)=0 \Rightarrow a_i(k)=a_i^{xy}(k),v_i(k)=v_i^{xy}(k)$ for notation convenience.} 
Secondly, we use the expectation maximization (EM) algorithm to obtain the most likely model parameters $\hat{\Theta}_i$ which fully define the distribution of the input vector $\mathbf{a}_i(k)$.  
This information is regarded as a noisy observation of the model parameters $\Theta_i$ and is fed into the Bayesian inference module in order to find the posterior probability of each class $c$ using prior distributions $P(c)$, and class conditional distributions $P(\Theta|c)=P(\Theta|\Psi_c)$ for all potential classes $c\in\{1,2,\dots,C\}$. The ultimate goal of this stage is to determine the most likely object class $c$ based on the observation $\hat{\Theta}_i$ as follows:
\begin{align} \label{eq:class}
c_i^{\star}= \underset{c=1,2,\dots,C}{\text{argmax}} P_c(c|\hat\Theta_i) 
\end{align}

In short, the most likely object class $c_i^{\star}$ is obtained by observing the object's motion trajectory $\mathbf{x}_i(k),k=0,1,2\dots,\infty$. In practice, a relatively short observation period (e.g. $k=0:1:100$) is sufficient for a reliable object motion profiling. Further, we use the statistical properties of the \textit{motion profiles} of the observed objects (i.e. $\Theta_i \in \mathcal{C}_{c_i}$) to refine the hyper parameters of each class, $P_c(\Psi)$ to further improve the prediction accuracy. In other words, we learn and refine class-specific mobility parameters using online observations, as detailed in the following section. Finally, we note that with our proposed flexible model, the system can admit new object types and the number of classes, $C$ in (\ref{eq:class}), can change over time to show the number of currently identified object classes.

%
%

\begin{figure}[b]
\begin{center}
\ifx \ncols \colone
\includegraphics[width=0.65\columnwidth]{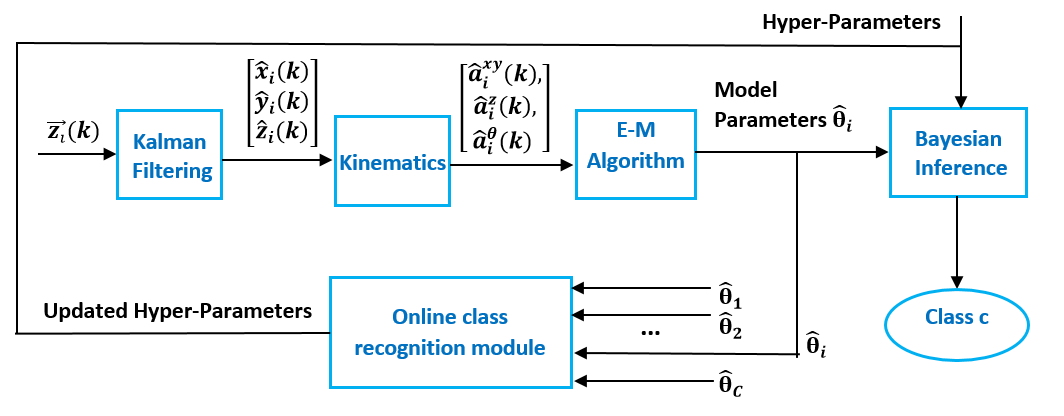}
\caption{Block diagram of the proposed joint mobility prediction and profiling (JMPP) method.}
\label{fig:method}
\end{center}
\vspace{-0.2 in}
\else
\includegraphics[width=0.8\columnwidth]{Fig4Method3D.png}
\caption{Block diagram of the proposed joint mobility prediction and profiling (JMPP) method.}
\label{fig:method}
\end{center}
\vspace{-0.4 in}
\fi
\end{figure}

\section{Joint Mobility Prediction and Profiling}   \label{sec:method} 
In this section, we elaborate on the details of the proposed method, which includes the following three steps. 

\textbf{Step 1- Denoising:} An accurate estimate of the state vector $\mathbf{x}_k$ can be obtained by solving the state transition equations in (\ref{eq:model}) using Kalman filtering, which includes two set of time update and measurement update equations. Time update equations are used to predict the next state vector (the location of the flying object in our modeling) based on the previous state. Measurement equations are used to refine the obtained prediction based on the noisy observation vector. The observation vector $\mathbf{z}_k$ represents the location information using an arbitrary tracking system. 
In the case of intermittent observation, the prediction is performed solely based on the time update equations, if no measurement is available. 
This system can be solved for an unknown input via optimal state estimation of singular systems~\cite{darouach1995kalman}, 
as shown in Fig. \ref{fig:kalman}. We use this approach to predict the future locations of a flying object by mitigating the system and measurement noise terms. Another commonly used approach is the prediction of the next location by linearizing the motion trajectory, which deems inefficient for highly non-linear motions. 
\begin{figure}[h]
\centering
\ifx \ncols \colone
\includegraphics[width=0.6\columnwidth]{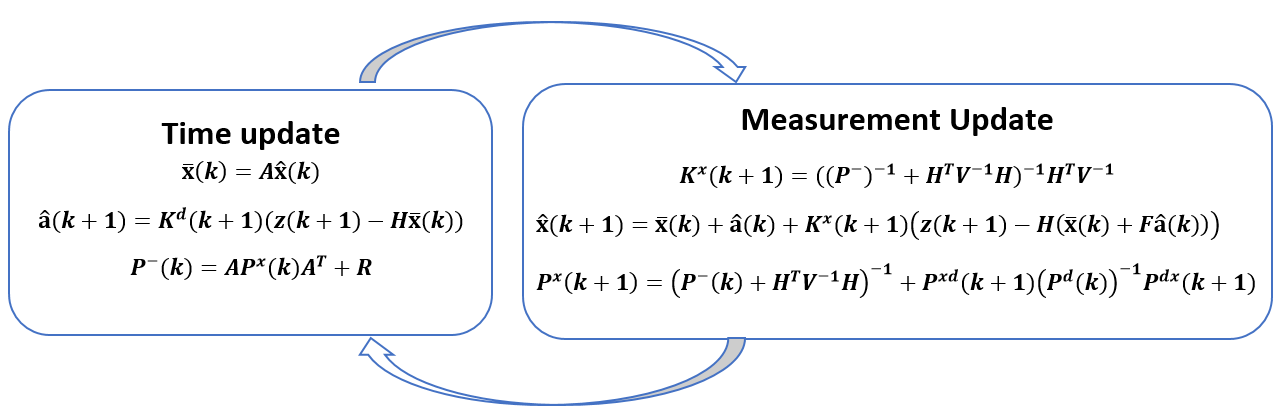} 
\caption{Kalman filtering with unknown input. In the absence of reliable measurement readings, prediction is performed solely based on time update equations and measurement update equations are skipped. Equations are from \cite{hsieh2010optimality}.}
\label{fig:kalman}
\else
\vspace{-0.2 in}
\includegraphics[width=.8\columnwidth]{Fig5KalmanEQ.png} 
\caption{Kalman filtering with unknown input. In the absence of reliable measurement readings, prediction is performed solely based on time update equations and measurement update equations are skipped. Equations are from \cite{hsieh2010optimality}.}
\label{fig:kalman}
\fi
\end{figure}

\textbf{Step 2- Mobility parameter extraction:}
The result of the previous stage provides an estimate of the object motion trajectory. The estimate of the instantaneous direct and angular accelerations can be readily obtained using kinematics equations, $\hat{a}_i(k)=\big(\hat{v}_i(k+1)-\hat{v}_i(k)\big)/dt$ and $\hat{a}_i^{\theta}(k)=\big(\hat{\theta}_i(k+1)-\hat{\theta}_i(k)\big)/dt$, where $\hat{v}_i(k)$ and $\hat{\theta}_i(k)$ are obtained from the estimated state vector $\hat{\mathbf{z}}_i(k)$ using (\ref{eq:v-w}).  
The acceleration parameters are modeled as a sequence of independent RVs with the defined distributions in (\ref{eq:acceleration}).  
If we model the estimation errors, $e_i(k)=\hat{a}_i(k)-a_i(k)$ and $e_i^{\theta}(k)=\hat{a}_i^{\theta}(k)-a_i^{\theta}(k)$, respectively with zero-mean Gaussian distributions of variance $\sigma_{n}^2$ and ${\sigma_{n}^{\theta}}^2$, then $\hat{a}_i(k)$ and $\hat{a}_i^{\theta}(k)$ for object $i$ follow Gaussian mixture model (GMM):
\begin{align}
\nonumber
&\hat{a}_i(k)\sim (1-\lambda_i)\mathcal{N}(0,\sigma_{n}^2)+  \lambda_i\mathcal{N}(\mu_i,\sigma_i^2+\sigma_{n}^2),\\
&\hat{a}_i^{\theta}(k) \sim (1-\lambda_i^{\theta})\mathcal{N}(0,{\sigma_{n}^{{\theta}}}^2)+ \lambda_i^{\theta}\mathcal{N}(\mu_i^{\theta},{\sigma_i^{{\theta}}}^2+{\sigma_{n}^{{\theta}}}^2),
\end{align}
where the model parameters $\Theta_i=(\lambda_i,\mu_i,\sigma_i^2,\lambda_i^{\theta},\mu_i^{\theta},{\sigma_i^{{\theta}}}^2)$ depend on the object class $c_i$ (represented by the vector of hyper-parameters $\Psi_{c_i}$). An optimal method to tune the parameters of a GMM based on its observations is the expectation maximization (EM) algorithm, which iterates between calculating the expected value of the likelihood function and maximizing the likelihood by updating membership probabilities. Here, we use EM to find the point estimates of $\Theta_i$, denoted by $\hat{\Theta}_i$ based on the observations $\hat{a}_i(k)$ and $\hat{a}_i^{\theta}(k)$, i.e.
\begin{align}
\hat{\Theta}_i = \underset{\Theta}{\text{argmax}}P(\Theta|\{\hat{a}_i(k),\hat{a}_i^{\theta}(k)\}_{k=0,1,2,\dots}).
\end{align}

\textbf{Step 3- Object profiling:} Note that the prior distribution of the model parameters $\Theta_i$, before observing the object's motion trajectory is $P(\Theta_i)=\sum_{c=1 }^C P(c)P(\Theta_i|c)=\sum_{c=1 }^C P(c)P(\Theta_i|\Psi_c)$. The output of EM algorithm for each segment of the motion trajectory is $\hat{\Theta}_i$, which can be considered as an observation of the actual $\Theta$. Therefore, we can use Bayes' rule to find the posterior probability of the objects class $c$ using 
\begin{align}
P(c|\hat{\Theta}_i)=\frac{P(\hat{\Theta}_i|\Psi_c)P(c)}{\sum_{c=1}^C P(\hat{\Theta}_i|\Psi_c)P(c)}
\end{align}
Here, we assign an equal probability for each class ($P(c)=1/C$ for $c=1,2,\dots,C$). Finally, the most likely class is determined using (\ref{eq:class}).

\begin{figure}[h]
\centering
\ifx \ncols \colone
\includegraphics[width=0.85\columnwidth]{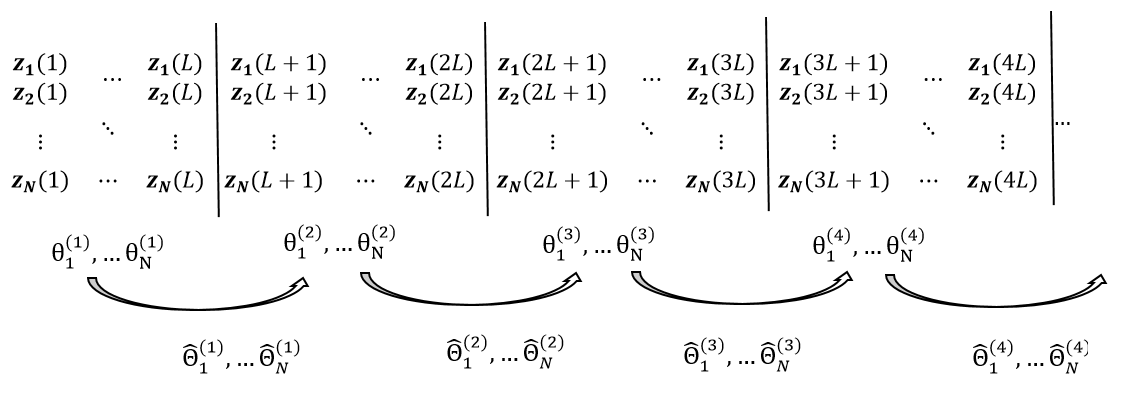} 
\else
\vspace{-0.2 in}
\includegraphics[width=0.85\columnwidth]{Fig6OnlineMethods.png} 
\fi
\caption{Online self class recognition module: motion profile of objects $\hat{\mathbf{\theta}}_1^{(s)}\dots\hat{\mathbf{\theta}}_N^{(s)}$ are found for all objects based on their observed trajectories $\mathbf{z}_i((s-1)L+1)\dots \mathbf{z}_i(sL)$ at segment $s$ as well as their previous updates.}
\label{fig:online}
\vspace{-0.2 in}
\end{figure}

\textbf{Step 4- Online class recognition module:} 
In the proposed algorithm as mentioned above, we considered a fixed number of motion classes with equal selection probabilities ($P(c)=1/C$). This may limit the applicability of the proposed method in practice due to the need for prior knowledge about the motion profile of each class represented by $\Psi_c$. Further, the system fails in addressing objects of new types with undefined motion profiles. In order to address this issue, we develop an online self tuning module. This module works based on segment wise processing of motion trajectories. Segment $s$ includes the observed locations during time interval $[(s-1)T,sT]$, namely $\mathbf{z}_i(k)$ for time steps $\{k=1,2,\dots| (s-1)T \leq  k ~\text{dt} \leq sT$ and targets $i=1,2,\dots,N$. Each segment includes $l=T/\text{dt}$ time points. 
At each segment $s$, we perform steps 1 and 2 to estimate the motion profile of each target node $i$ and we show it with $\hat{\mathbf{\theta}}_i^{(s)}$. Then, we obtain the current motion profile of object $i$, $\hat{\Theta}_i^{(s)}$ using: 
\ifx \ncols \coltwo
\vspace{-0.1 in}
\begin{align}
\hat{\Theta}_i^{(s)}=\big[(s-1)\hat{\Theta}_i^{(s-1)}+{\hat{\mathbf{\theta}}}_i^{(s)}\big]/s
\end{align}
\else
\begin{align}
\hat{\Theta}_i^{(s)}=\frac{(s-1)\hat{\Theta}_i^{(s-1)}+{\hat{\mathbf{\theta}}}_i^{(s)}}{s}
\end{align}
\fi 
Note that we have $\hat{\Theta}_i^{(s)}=\frac{\sum_{t=1}^s \hat{\mathbf{\theta}}_i^{(s)}}{s}$. With this online method, we use the previous estimate of the motion profiles $\hat{\Theta}_i^{(s-1)}$ and only process the last segment of the received trajectory to obtain $\hat{\Theta}_i^{(s)}$, which is more efficient than processing the entire history of the trajectory. Then, we proceed with step 3 and find the most likely class of each target based on $\hat{\Theta}_i^{(s)}$, and the most recent estimate of the hyper-parameters $\hat{\Psi}_c^{(s)}$ for $c=1,2,\dots,C^{(s)}$. The main distinction here is that the number of classes and their representative hyper-parameters $\hat{\Psi}_c^{(s)}$ are not fixed anymore and they rather are learned from the observed trajectories as depicted in figure \ref{fig:online}. 

Once we complete steps 1-3 for all objects $i=1,2,\dots,N$ during interval $[(s-1)T,sT]$, we cluster the collected motion profiles, represented by vectors $\hat{\Theta}_i^{(s)}$. We use parametric clustering, and try the number of clusters $n_C$ within the $[C^{(s-1)}-\Delta C,C^{(s-1)}+ \Delta C]$ range, where $C^{(s-1)}$ is the number of previously recognized clusters, and $\pm \Delta C$ enables genesis of new clusters as well as death of fake clusters. Therefore, the number of valid motion classes can change over time if a reasonable evidence is provided by the accumulated motion trajectories. In order to identify the optimal number of clusters, we consider within-cluster variance penalized by the number of clusters using $f(\mathcal{A}_{n_C})=\alpha \Sigma_{\text{w}}+(1-\alpha)\beta n_C$. Here, $\mathcal{A}_{n_C}$ is the clustering algorithm with $n_C$ clusters, and $\Sigma_{\text{w}}$ is the resulting averaged within-variance defined as
\begin{align}
\ifx \ncols \coltwo
\nonumber
\fi
&\Sigma_{\text{w}}=\frac{1}{{n_C(|C_1|+\dots+|C_{n_C}|)}}\sum_{c=1}^{n_C} \sum_{\hat{\Theta}_i \in C_c}(\hat{\Theta}_i-\mu_c)^2, 
\ifx \ncols \coltwo
\\
\fi
&\mu_c = \frac{1}{{|C_c|}} \sum_{\hat{\Theta}_i \in C_c}\hat{\Theta}_i,
\end{align}
where $|C|$ is the number of elements in set $C$. In the simulation results in section \ref{sec:simulation}, we use K-means clustering and set $C^{(1)}=4, \Delta C =3$, $\alpha=0.2$ and $\beta=10$ using cross-validation. The number of clusters is obtained as
\begin{align}
C^{(s)}=\underset{n_C \in [C^{(s-1)}-\Delta C,\dots,C^{(s-1)}+ \Delta C]}{\text{argmin}}{f(\mathcal{C}_n)}
\end{align}

Each cluster represents a mobility class $c=1,2,\dots,n_C$. Therefore, the collected motion profiles of each cluster ($\{ \hat\Theta_i: i \in \text{cluster }c \}$) is used to refine the relevant clusters hyper parameters $\Psi_c$ by applying maximum likelihood estimation (MLE) to (\ref{eq:prior}).

\ifx \ncols \coltwo
\vspace{-0.2 in}
\fi
\section{Simulation Results} \label{sec:simulation} 
In this section, simulation results are provided to assess the performance of the proposed method in comparison with the state of the art. \textcolor{black}{Here, we assume that each drone is equipped with a tracking system and hence can monitor and estimate the location of surrounding objects. For instance, Lidar systems, ultrasound systems, or visual cameras can be used to accurately measure the surrounding objects \cite{razi2018predictive}. However, most off-the shelf commercial drones (e.g. DJI phantom and Matrice series) do not include pricey tracking systems. For such scenarios, ADS-B technology \cite{drone-ADS-B} can be used where drones locate themselves using embedded GPS positioning modules and periodically propagate their positions to other nodes according, to be used for trajectory prediction. For drones in an adversary network, a ground-based tracking system (e.g. a conventional Radar) can be used to locate the flying objects and perform the object classification task.}

We use the following simulation parameters unless otherwise specified. We define $C=3$ clusters with hyper-parameters $\Psi_1,\Psi_2,\Psi_3$ shown in Table \ref{tab:1}. We use state transition and measurement equations in (\ref{eq:model}) to develop random motion trajectories as well as their linear measurements for $N=300$ objects ($100$ per class). The system and measurement noise variances are set to \textcolor{black}{$\mathbf{R}=\mathbf{I}_{6\times6}$ and $\mathbf{Q}=\mathbf{I}_{3\times3}$.} 

\begin{table}[h]
\centering
\caption{Motion profiles for three classes ($\Psi_1,\Psi_2,\Psi_3$).}
\begin{tabular}{|c | c| c| c|c|c|c|}
\hline
\multirow{2}{*}{class} &

\multicolumn{3}{c|}{Speed hyper-parameters}& \multicolumn{3}{c|}{Direction hyper-parameters} \\
\cline{2-7}
& $(a_1,b_1)$ & $(\alpha_1,\beta_1)$& $n_{1}$ & $(a_1,b_1)$ & $(\alpha_1,\beta_1)$&  $n_{2}$ \\ 
\hline
 1 & (2,50) & (10,10) & 1 &(2,50)&(10,10) & 1\\ 
 2 & (4,4) & (2,0.5) & 2 &(4,4)&(2,0.5) & 2\\ 
 3 & (50,2) & (2,0.1) & 10 &(50,2)&(2,0.1) & 10\\ 
\hline
\end{tabular}
\label{tab:1}
\end{table}

\begin{figure}[t]
\centering
\ifx \ncols \colone
\includegraphics[width=0.5\columnwidth]{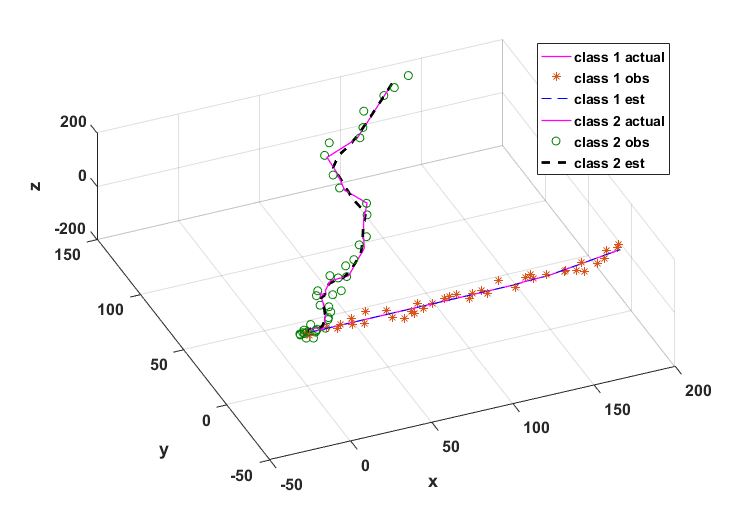}
\else
\includegraphics[width=.85\columnwidth]{Fig7Trajectory.png}
\fi
\caption{\textcolor{black}{Estimating motion trajectories of two objects of different types in 3D space with unknown input (driving force) vector.}}
\label{fig:kalman-2class}
\vspace{-0.2 in}
\end{figure}

The results of the first stage using Kalman filtering with unknown input are presented in Fig. \ref{fig:kalman-2class} for two objects belonging to different mobility classes ($C_1$ and $C_2$). The results show a relatively accurate estimation of locations provided by step-1 of the proposed method for further analysis. The mean squared error ratio $\sum(\mathbf x_i-\hat{\mathbf  x_i})^2/\sum \mathbf x_i^2$ is less than $1\%$ for both classes.

\begin{figure}[h]
\centering
\ifx \ncols \colone
	\includegraphics[width=0.99\columnwidth]{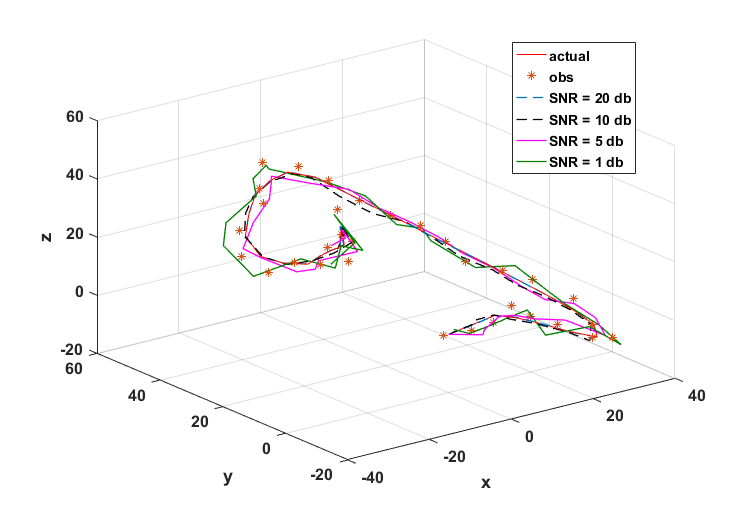}
\caption{\textcolor{black}{The impact of measurement update rate ($r$) on the accuracy of motion trajectory prediction.}}
	\label{fig:kalman-intermittent}
\else
	\includegraphics[width=0.7\columnwidth]{Fig8KFvsSNR.png}
\caption{\textcolor{black}{The impact of measurement update rate ($r$) on the accuracy of motion trajectory prediction.}}
	\label{fig:kalman-intermittent}
	\vspace{-0.2 in}
\fi
\end{figure}


%
\textcolor{black}{Fig. \ref{fig:kalman-intermittent} investigates the impact of measurement update rate ($r$) on the mobility prediction for the case of intermittent observations. This parameter determines the probability of successful observation attempts, where the value of $\mathbf{z}_i(k)$ is valid.} This case is more important and shows the utility of the proposed method in predicting future node positions, when the measurement readings are not available. The prediction accuracy significantly declines if the measurement update rate ($r$) goes below an acceptable level. 

The second utility of the proposed method is object profiling based an motion trajectories. The results of motion profiling accuracy are presented in Table \ref{tab:profiling} for $N=300$ randomly generate motions trajectories. \textcolor{black}{The results are promising and exhibit an average classification success rate (CSR) of $(90+91+92)/(100+100+100)\approx 91\%$.} These results verify the success of three sequential steps in jointly predicting the motion trajectories and profiling the objects into correct mobility classes. As shown in Fig. \ref{fig:snr}, this accuracy depends on the quality of the trajectory estimation, which in turn is influenced by the tracking system noise level.

There are very few prior works that consider profiling object classes based on their online motion trajectories. The most closest work we found is \cite{dodge2009revealing}, which proposes a method to classify moving point objects (MPO) based on their motion patterns. This method, we call it MPO, is based on extracting straightness and velocity indexes from the motion trajectories. Further, they classified objects such as cards, pedestrians, bicycles, and motorcycles based on statistical features (e.g. mean, median, min, max, skewness and standard deviation) of the mobility indexes. Here, we compare our method against this method. We also applied common classification methods such as fuzzy c-means (FCM), and K-means directly to the datapoints of the estimated driving forces ($\mathbf{a}_i(k),\mathbf{a}_i^\theta(k)$) for each trajectory for the sake of completeness. Finally, inspired by other time-series analysis (e..g ECG signal processing), we trained a Gaussian process (GP) for the observed trajectories to exploit the fundamental property of each trajectory and then classified the objects based on the obtained GP parameters. The results of this comparison are provided in Table \ref{tab:Comparison} for 300 objects whose trajectories are simulated using three different classes. The comparison shows that our method (JMPP) overcomes all methods by a significant margin, since the proposed method tries to directly recover motion profiles in a reverse-engineering fashion. \textcolor{black}{The proposed method achieves a CSR of $91\%$ compared to $80.33\%$ obtained using GP.} 


\begin{table}[t]
\centering
{\color{black}
\caption{\textcolor{black}{Motion profiling accuracy of the proposed method in 3D space.}}
\begin{tabular}{ |c|c|c|c|}
 \hline
Actual Class &  \multirow{2}{*}{C1}&  \multirow{2}{*}{C2}&  \multirow{2}{*}{C3} \\
\cline{1-1}
Predicted Class & & &\\
\hline
 C1   & 90   & 6   &  4\\
 C2   & 5    & 91  &  4\\
 C3   & 6    & 2   & 92\\
 \hline
\end{tabular}
\label{tab:profiling}
}
\end{table}

\begin{table}[t]
\centering
{\color{black}
\caption{\textcolor{black}{Classification success rate of different object profiling methods based on 3D motion trajectories.}}
\begin{tabular}{|p{0.75 cm}|c|c|c|c|c|c|}
 \hline
\multirow{2}{*}{Class} & \multirow{2}{*} {\# of Traj. }& \multicolumn{5}{|c|}{Number of Correctly Classified Object}   \\
\cline{3-7}
&&{K-mean}&FCM&  {MPO} & {GP} & {JMPP}\\
\hline
 C1  & 100 & 36 & 41   & 79   & 81  &  90\\
 C2  & 100 & 44 & 46   & 84   & 85  &  91\\
 C3  & 100 & 35 & 37   & 80   & 84  &  92\\
 \hline
 Total &300 &115 &124 &243 & 250 & 273\\
 \hline
CSR & $100\%$ & $38.33\%$ & $41.33\%$ &$81.0\%$ &$80.33\%$& $91\%$\\
 \hline
\end{tabular}
\label{tab:Comparison}
\ifx \ncols \coltwo
\vspace{-0.2 in}
\fi
}\end{table}

\begin{figure}[t]
\centering
\ifx \ncols \colone
\includegraphics[width=0.5\columnwidth]{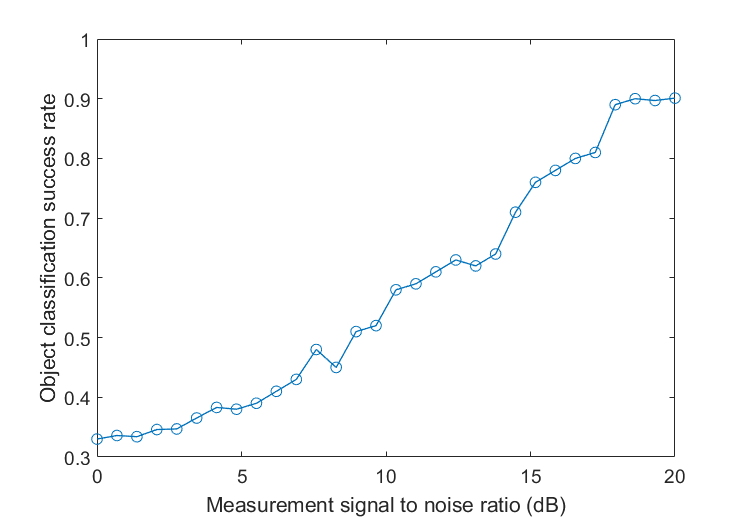}
\caption{Classification success rate for object profiling based on 2D motion patterns versus signal to measurement noise level.}
\label{fig:snr}
\else
\includegraphics[width=0.7\columnwidth]{Fig9CSRvsSNR.png}
\caption{Classification success rate for object profiling based on 2D motion patterns versus signal to measurement noise level.}
\label{fig:snr}
\vspace{-0.2 in}
\fi

\end{figure}

Finally, we investigate the performance of the online class recognition method. This module works based on clustering motion profile vectors with a penalized number of clusters. Two key features of this method are online-learning of class-specific hyper-parameters as well as recognizing new objects as they enter the system. These two properties are illustrated in Figs. \ref{fig:Online self-tuning result} and \ref{fig:cluster}, respectively. Fig. \ref{fig:Online self-tuning result} investigates the accuracy of class-specific model hyper-parameters in comparison with the actual ones used to generate the motion trajectories. The accuracy is represented in terms of mean squared errors (MSE) ratio. For instance if vectors $\Psi$ and $\Psi^\prime$, represent the actual and estimate vector of hyper-parameters for all classes, the MSE is calculated as $|\Psi-\Psi^\prime|_2^2/|\Psi|_2^2$, where $|.|_2$ is the second norm.
The results show that the MSE error remains within $20\%$ after receiving a few trajectory segments. However, the performance also depends on the length of each segment. The results show that longer trajectory segments provide more accurate estimate of hyper-parameters. For instance, using segment length of $80$ points, ensures that the MSE error remains below $3\%$ after receiving as few as 10 segments. Therefore, the system does not need to have prior knowledge about the motion properties of different object classes, which makes it more desirable for practical situations.

\begin{figure}[t]
\centering
\ifx \ncols \colone
\includegraphics[width=0.5\columnwidth]{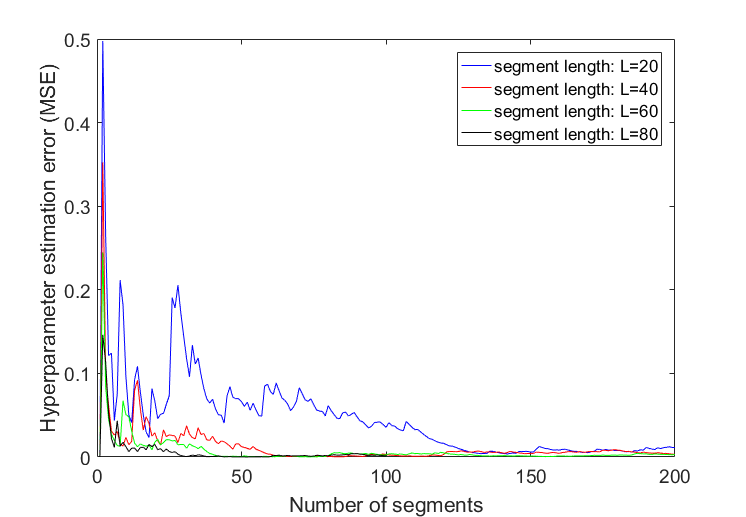}
\caption{The performance of the online self tuning module in estimating class-specific hyper-parameters from the observed motion trajectories in terms of mean squared errors (MSE).}
\label{fig:Online self-tuning result}
\else
\includegraphics[width=0.7\columnwidth]{Fig10MSEvsSegLen.png}
\caption{The performance of the online self tuning module in estimating class-specific hyper-parameters from the observed motion trajectories in terms of mean squared errors (MSE).}
\label{fig:Online self-tuning result}
\vspace{-0.2 in}
\fi
\end{figure}

Fig. \ref{fig:cluster} illustrates the capability of the system to recognize and profile new objects with unseen motion properties.
For this part, we start with $C=4$ clusters and generate $N=100$ objects for each class. The system processes the observed motion trajectories and correctly recognizes $C=4$ clusters. Now, we start adding objects of a new type with an unseen motion profile to the system. The system, after collecting a few new objects, recognizes the existence of new object class and changes the number of clusters to $C=4+1$. We repeat this experiment $100$ times and define the probability of correctly recognizing new classes after receiving $n$ objects as the ratio of the number of experiments that reports $C=5$ (after observing $n$ new objects) to the total number of experiments. The results are shown in Fig. \ref{fig:cluster}. For instance, the system recognizes the arrival of new object type with probability $80\%$ after receiving $n=44$ objects of this new type. The accuracy approaches $100\%$ after receiving about $55$ objects of the new type. Therefore, this module enables the system to adaptively generate new object classes over time in addition to tuning the hyper-parameters of existing classes.

\begin{figure}[h]
\centering
\ifx \ncols \colone
\includegraphics[width=0.5\columnwidth]{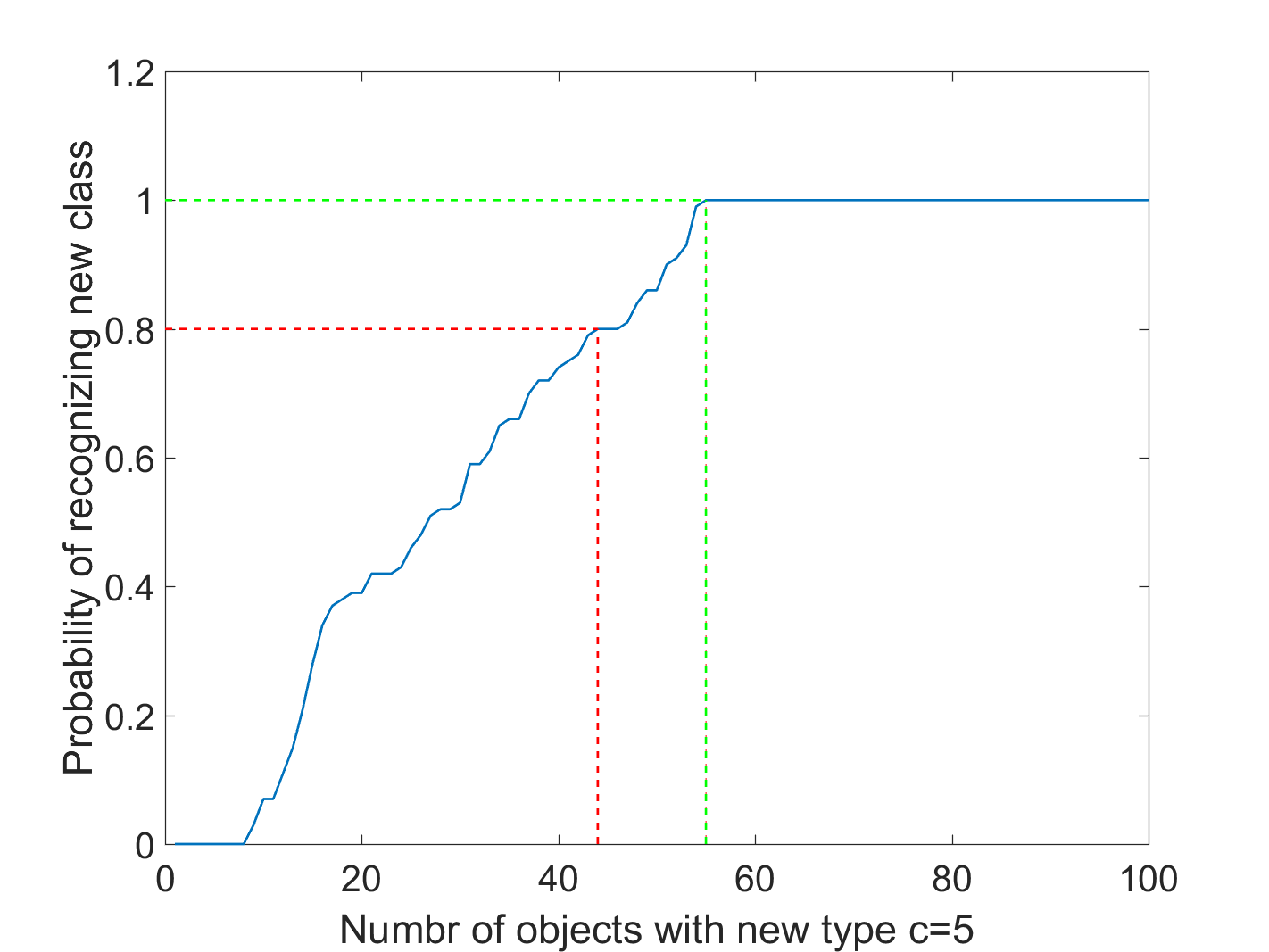}
\caption{The performance of online self tuning module. The probability of successfully determining the genesis of new classes represented versus the number of objects with new motion profiles in 2D space ($v_z=0$) averaged over 100 runs.}
\label{fig:cluster}
\else
\includegraphics[width=0.7\columnwidth]{Fig11NewClass.png}
\caption{The performance of online self tuning module. The probability of successfully determining the genesis of new classes represented versus the number of objects with new motion profiles in 2D space ($v_z=0$) averaged over 100 runs.}
\label{fig:cluster}
\vspace{-0.2 in}
\fi
\end{figure}

\ifx \ncols \coltwo
\vspace{-0.2 in}
\fi
\section{Conclusions} \label{sec:conclusions}
In this work, a novel framework is proposed for joint mobility prediction and profiling of objects through analyzing their motion trajectories. The idea is to process the motion trajectories in terms of state transition equations to predict the objects future locations and extract the driving forces. Also, we develop a natural hierarchical generative model for the exerted direct and rotational forces. This approach enables us to exploit the motion properties of mobile objects and classify them based on their motion properties. 
Compared to other methods, our unified framework neither requires a large training dataset (as opposed to data-driven methods) nor is tailored to a specific object class (as opposed to model-based methods). The proposed method yields a success rate of $90\%$ in profiling mobile objects for a reasonable measurement noise level which shows $7\%$ improvement compared to the state of the art method. 

Further, a novel online self-tuning algorithm is proposed which tunes the general motion properties of each class (represented by the class-specific hyper-parameters) by processing the accumulated trajectories over time. This approach adaptively generates new motion classes by observing objects with unseen motion profiles. Therefore, no prior information is required about the motion dynamics of different object types, which makes this system desirable for practical applications.
The proposed algorithm, if integrated with communication protocols (e.g. routing algorithms in network layer), can facilitate information flow in UAV networks and IoT with flying objects by predicting the future network topology.  

\ifx \ncols \coltwo
\vspace{-0.1 in}
\fi
\section{ACKNOWLEDGMENT}
This material is based upon work supported by the National Science Foundation under Grant No. 1755984.
The authors also acknowledge the U.S. Government's support in the publication of this paper. This material is based upon work
partially funded by AFRL, under AFRL Grant No. FA8075-14-D-0014. Any opinions, findings and conclusions or recommendations
expressed in this material are those of the author(s) and do not necessarily reflect the views of the US government or AFRL.

\bibliographystyle{IEEEtran}   
\bibliography{BibHan,RzBibJCN}

\begin{thebibliography}{10}
\providecommand{\url}[1]{#1}
\csname url@samestyle\endcsname
\providecommand{\newblock}{\relax}
\providecommand{\bibinfo}[2]{#2}
\providecommand{\BIBentrySTDinterwordspacing}{\spaceskip=0pt\relax}
\providecommand{\BIBentryALTinterwordstretchfactor}{4}
\providecommand{\BIBentryALTinterwordspacing}{\spaceskip=\fontdimen2\font plus
\BIBentryALTinterwordstretchfactor\fontdimen3\font minus
  \fontdimen4\font\relax}
\providecommand{\BIBforeignlanguage}[2]{{%
\expandafter\ifx\csname l@#1\endcsname\relax
\typeout{** WARNING: IEEEtran.bst: No hyphenation pattern has been}%
\typeout{** loaded for the language `#1'. Using the pattern for}%
\typeout{** the default language instead.}%
\else
\language=\csname l@#1\endcsname
\fi
#2}}
\providecommand{\BIBdecl}{\relax}
\BIBdecl

\bibitem{thiels2015use}
C.~A. Thiels, J.~M. Aho, S.~P. Zietlow, and D.~H. Jenkins, ``Use of unmanned
  aerial vehicles for medical product transport,'' \emph{Air medical journal},
  vol.~34, no.~2, pp. 104--108, 2015.

\bibitem{kanistras2015survey}
K.~Kanistras, G.~Martins, M.~J. Rutherford, and K.~P. Valavanis, ``Survey of
  unmanned aerial vehicles (\textsc{UAV}s) for traffic monitoring,'' in
  \emph{Handbook of unmanned aerial vehicles}.\hskip 1em plus 0.5em minus
  0.4em\relax Springer, 2015, pp. 2643--2666.

\bibitem{everaerts2008use}
J.~Everaerts \emph{et~al.}, ``The use of unmanned aerial vehicles
  (\textsc{UAV}s) for remote sensing and mapping,'' \emph{The International
  Archives of the Photogrammetry, Remote Sensing and Spatial Information
  Sciences}, vol.~37, no. 2008, pp. 1187--1192, 2008.

\bibitem{xu2016internet}
J.~Xu, G.~Solmaz, R.~Rahmatizadeh, D.~Turgut, and L.~Boloni, ``Internet of
  things applications: animal monitoring with unmanned aerial vehicle,''
  \emph{arXiv preprint arXiv:1610.05287}, 2016.

\bibitem{freeman2015agricultural}
P.~K. Freeman and R.~S. Freeland, ``Agricultural \textsc{UAV}s in the us:
  potential, policy, and hype,'' \emph{Remote Sensing Applications: Society and
  Environment}, vol.~2, pp. 35--43, 2015.

\bibitem{wall2011surveillance}
T.~Wall and T.~Monahan, ``Surveillance and violence from afar: The politics of
  drones and liminal security-spaces,'' \emph{Theoretical Criminology},
  vol.~15, no.~3, pp. 239--254, 2011.

\bibitem{joo2018low}
C.~Joo and J.~Choi, ``Low-delay broadband satellite communications with
  high-altitude unmanned aerial vehicles,'' \emph{Journal of Communications and
  Networks}, vol.~20, no.~1, pp. 102--108, 2018.

\bibitem{girard2004border}
A.~R. Girard, A.~S. Howell, and J.~K. Hedrick, ``Border patrol and surveillance
  missions using multiple unmanned air vehicles,'' in \emph{Decision and
  Control, 2004. CDC. 43rd IEEE Conference on}, vol.~1.\hskip 1em plus 0.5em
  minus 0.4em\relax IEEE, 2004, pp. 620--625.

\bibitem{faa_2016}
M.~Huerta, ``Drones: A story of revolution and evolution,'' Jan 2017.

\bibitem{gupta2016survey}
L.~Gupta, R.~Jain, and G.~Vaszkun, ``Survey of important issues in \textsc{UAV}
  communication networks,'' \emph{IEEE Communications Surveys \& Tutorials},
  vol.~18, no.~2, pp. 1123--1152, 2016.

\bibitem{quaritsch2010networked}
M.~Quaritsch, K.~Kruggl, D.~Wischounig-Strucl, S.~Bhattacharya, M.~Shah, and
  B.~Rinner, ``Networked uavs as aerial sensor network for disaster management
  applications,'' \emph{e \& i Elektrotechnik und Informationstechnik}, vol.
  127, no.~3, pp. 56--63, 2010.

\bibitem{rosati2016dynamic}
S.~Rosati, K.~Kru{\.z}elecki, G.~Heitz, D.~Floreano, and B.~Rimoldi, ``Dynamic
  routing for flying ad hoc networks,'' \emph{IEEE Transactions on Vehicular
  Technology}, vol.~65, no.~3, pp. 1690--1700, 2016.

\bibitem{kaleem2018amateur}
Z.~Kaleem and M.~H. Rehmani, ``Amateur drone monitoring: State-of-the-art
  architectures, key enabling technologies, and future research directions,''
  \emph{IEEE Wireless Communications}, vol.~25, no.~2, pp. 150--159, 2018.

\bibitem{Afghah_ACC18}
\BIBentryALTinterwordspacing
F.~Afghah, M.~Zaeri{-}Amirani, A.~Razi, J.~Chakareski, and E.~S. Bentley, ``A
  coalition formation approach to coordinated task allocation in heterogeneous
  {UAV} networks,'' \emph{CoRR}, vol. abs/1711.00214, 2017. [Online].
  Available: \url{http://arxiv.org/abs/1711.00214}
\BIBentrySTDinterwordspacing

\bibitem{Razi_Asilomar17}
A.~Razi, F.~Afghah, and J.~Chakareski, ``Optimal measurement policy for
  predicting \textsc{UAV} network topology,'' in \emph{51th Asilomar Conference
  on Signals, Systems and Computers (Asilomar'17)}, 2017.

\bibitem{7820569}
A.~Anand, H.~Aggarwal, and R.~Rani, ``Partially distributed dynamic model for
  secure and reliable routing in mobile ad hoc networks,'' \emph{Journal of
  Communications and Networks}, vol.~18, no.~6, pp. 938--947, Dec 2016.

\bibitem{6182876}
V.~Sharma, K.~Kar, R.~La, and L.~Tassiulas, ``Dynamic network provisioning for
  time-varying traffic,'' \emph{Journal of Communications and Networks},
  vol.~9, no.~4, pp. 408--418, Dec 2007.

\bibitem{6182872}
A.~Urra, E.~Calle, J.~L. Marzo, and P.~Vila, ``An enhanced dynamic multilayer
  routing for networks with protection requirements,'' \emph{Journal of
  Communications and Networks}, vol.~9, no.~4, pp. 377--382, Dec 2007.

\bibitem{zhang2014hdre}
Y.~Zhang, X.~Zhang, W.~Fu, Z.~Wang, and H.~Liu, ``Hdre: Coverage hole detection
  with residual energy in wireless sensor networks,'' \emph{Journal of
  Communications and Networks}, vol.~16, no.~5, pp. 493--501, 2014.

\bibitem{linkselection}
J.~H. Sarker and R.~Jantti, ``Connectivity modeling of wireless multihop
  networks with correlated and independent factors,'' in \emph{The 6th
  International Conference on Advanced Communication Technology, 2004.},
  vol.~1, Feb 2004, pp. 474--479.

\bibitem{khaledi2018greedy}
M.~Khaledi, A.~Rovira-Sugranes, F.~Afghah, and A.~Razi, ``On greedy routing in
  dynamic \textsc{UAV} networks,'' \emph{arXiv preprint arXiv:1806.04587},
  2018.

\bibitem{ArnauWiSEE}
A.~Sugranes and A.~Razi, ``Predictive routing for dynamic \textsc{UAV}
  networks,'' in \emph{IEEE International Conference on Wireless for Space and
  Extreme Environments (WiSEE)}, Oct 2017.

\bibitem{Data_driven}
M.~Heß, F.~Büther, and K.~P. Schäfers, ``Data-driven methods for the
  determination of anterior-posterior motion in pet,'' \emph{IEEE Transactions
  on Medical Imaging}, vol.~36, no.~2, pp. 422--432, Feb 2017.

\bibitem{TAPASCologne1}
\BIBentryALTinterwordspacing
``Vehicular mobility trace of the city of cologne, germany,'' 2016. [Online].
  Available: \url{http://kolntrace.project.citi-lab.fr/}
\BIBentrySTDinterwordspacing

\bibitem{Model-based}
R.~J. Schalkoff and X.~Wang, ``A model-based viewpoint determination method for
  multiple object 3-d motion estimation,'' in \emph{Southeastcon '89.
  Proceedings. Energy and Information Technologies in the Southeast., IEEE},
  Apr 1989, pp. 1074--1079 vol.3.

\bibitem{choi2006learning}
P.~P. Choi and M.~Hebert, ``Learning and predicting moving object trajectory: a
  piecewise trajectory segment approach,'' \emph{Robotics Institute}, p. 337,
  2006.

\bibitem{bennewitz2005learning}
M.~Bennewitz, W.~Burgard, G.~Cielniak, and S.~Thrun, ``Learning motion patterns
  of people for compliant robot motion,'' \emph{The International Journal of
  Robotics Research}, vol.~24, no.~1, pp. 31--48, 2005.

\bibitem{gonzalez2008understanding}
M.~C. Gonzalez, C.~A. Hidalgo, and A.-L. Barabasi, ``Understanding individual
  human mobility patterns,'' \emph{Nature}, vol. 453, no. 7196, pp. 779--782,
  2008.

\bibitem{aoude2011mobile}
G.~Aoude, J.~Joseph, N.~Roy, and J.~How, ``Mobile agent trajectory prediction
  using bayesian nonparametric reachability trees,'' in \emph{Infotech@
  Aerospace 2011}, 2011, p. 1512.

\bibitem{lee2012identification}
G.~Lee, R.~Mallipeddi, and M.~Lee, ``Identification of moving vehicle
  trajectory using manifold learning,'' in \emph{International Conference on
  Neural Information Processing}.\hskip 1em plus 0.5em minus 0.4em\relax
  Springer, 2012, pp. 188--195.

\bibitem{human}
E.~Malmi, ``Human mobility prediction: A probabilistic transfer learning
  approach,'' in \emph{Aalto University School of Science}, Feb 2013.

\bibitem{self-propellers}
A.~Nourhani, P.~Lammert, A.~Borhan, and V.~Crespi, ``Kinematic matrix theory
  and universalities in self-propellers and active swimmers,'' in \emph{Phys
  Rev E Stat Nonlin Soft Matter Phys}, Jun 2014.

\bibitem{rovers}
M.~Tarokh and G.~J. McDermott, ``Kinematics modeling and analyses of
  articulated rovers,'' \emph{IEEE Transactions on Robotics}, vol.~21, no.~4,
  pp. 539--553, Aug 2005.

\bibitem{jia2013target}
X.~Jia, Z.~L. Wu, and H.~Guan, ``The target vehicle movement state estimation
  method with radar based on kalman filtering algorithm,'' in \emph{Applied
  Mechanics and Materials}, vol. 347.\hskip 1em plus 0.5em minus 0.4em\relax
  Trans Tech Publ, 2013, pp. 638--642.

\bibitem{sharma2016trajectory}
R.~Sharma and T.~Guha, ``A trajectory clustering approach to crowd flow
  segmentation in videos,'' in \emph{Image Processing (ICIP), 2016 IEEE
  International Conference on}.\hskip 1em plus 0.5em minus 0.4em\relax IEEE,
  2016, pp. 1200--1204.

\bibitem{gariel2011trajectory}
M.~Gariel, A.~N. Srivastava, and E.~Feron, ``Trajectory clustering and an
  application to airspace monitoring,'' \emph{IEEE Transactions on Intelligent
  Transportation Systems}, vol.~12, no.~4, pp. 1511--1524, 2011.

\bibitem{yuan2012efficient}
G.~Yuan, S.~Xia, L.~Zhang, Y.~Zhou, and C.~Ji, ``An efficient
  trajectory-clustering algorithm based on an index tree,'' \emph{Transactions
  of the Institute of Measurement and Control}, vol.~34, no.~7, pp. 850--861,
  2012.

\bibitem{mao2017adaptive}
Y.~Mao, H.~Zhong, H.~Qi, P.~Ping, and X.~Li, ``An adaptive trajectory
  clustering method based on grid and density in mobile pattern analysis,''
  \emph{Sensors}, vol.~17, no.~9, p. 2013, 2017.

\bibitem{xu2015unsupervised}
H.~Xu, Y.~Zhou, W.~Lin, and H.~Zha, ``Unsupervised trajectory clustering via
  adaptive multi-kernel-based shrinkage,'' in \emph{Proceedings of the IEEE
  International Conference on Computer Vision}, 2015, pp. 4328--4336.

\bibitem{yuan2017review}
G.~Yuan, P.~Sun, J.~Zhao, D.~Li, and C.~Wang, ``A review of moving object
  trajectory clustering algorithms,'' \emph{Artificial Intelligence Review},
  vol.~47, no.~1, pp. 123--144, 2017.

\bibitem{mcfadyen2016aircraft}
A.~Mcfadyen, M.~O'Flynn, T.~Martin, and D.~Campbell, ``Aircraft trajectory
  clustering techniques using circular statistics,'' in \emph{Aerospace
  Conference, 2016 IEEE}.\hskip 1em plus 0.5em minus 0.4em\relax IEEE, 2016,
  pp. 1--10.

\bibitem{pires2017shape}
T.~J. Pires and M.~A. Figueiredo, ``Shape-based trajectory clustering.'' in
  \emph{ICPRAM}, 2017, pp. 71--81.

\bibitem{valehi2017maximizing}
A.~Valehi and A.~Razi, ``Maximizing energy efficiency of cognitive wireless
  sensor networks with constrained age of information,'' \emph{IEEE
  Transactions on Cognitive Communications and Networking}, vol.~3, no.~4, pp.
  643--654, 2017.

\bibitem{valehi2018online}
------, ``An online learning method to maximize energy efficiency of cognitive
  sensor networks,'' \emph{IEEE Communications Letters}, vol.~22, no.~5, pp.
  1050--1053, 2018.

\bibitem{jiang2017machine}
C.~Jiang, H.~Zhang, Y.~Ren, Z.~Han, K.-C. Chen, and L.~Hanzo, ``Machine
  learning paradigms for next-generation wireless networks,'' \emph{IEEE
  Wireless Communications}, vol.~24, no.~2, pp. 98--105, 2017.

\bibitem{alsheikh2014machine}
M.~A. Alsheikh, S.~Lin, D.~Niyato, and H.-P. Tan, ``Machine learning in
  wireless sensor networks: Algorithms, strategies, and applications,''
  \emph{IEEE Communications Surveys \& Tutorials}, vol.~16, no.~4, pp.
  1996--2018, 2014.

\bibitem{xie2013comprehensive}
J.~Xie, Y.~Wan, K.~Namuduri, S.~Fu, and J.~Kim, ``A comprehensive modeling
  framework for airborne mobility,'' in \emph{AIAA Infotech@ Aerospace (I@ A)
  Conference}, 2013, p. 5053.

\bibitem{fotouhi2017dronecells}
A.~Fotouhi, M.~Ding, and M.~Hassan, ``Dronecells: Improving 5g spectral
  efficiency using drone-mounted flying base stations,'' \emph{arXiv preprint
  arXiv:1707.02041}, 2017.

\bibitem{wang2010novel}
W.~Wang, X.~Guan, B.~Wang, and Y.~Wang, ``A novel mobility model based on
  semi-random circular movement in mobile ad hoc networks,'' \emph{Information
  Sciences}, vol. 180, no.~3, pp. 399--413, 2010.

\bibitem{bouachir2014mobility}
O.~Bouachir, A.~Abrassart, F.~Garcia, and N.~Larrieu, ``A mobility model for
  uav ad hoc network,'' in \emph{Unmanned Aircraft Systems (ICUAS), 2014
  International Conference on}.\hskip 1em plus 0.5em minus 0.4em\relax IEEE,
  2014, pp. 383--388.

\bibitem{BiShop-Book}
C.~Bishop, ``Pattern recognition and machine learning,'' Jan 2006, p. 117.

\bibitem{darouach1995kalman}
M.~Darouach, M.~Zasadzinski, A.~B. Onana, and S.~Nowakowski, ``Kalman filtering
  with unknown inputs via optimal state estimation of singular systems,''
  \emph{International journal of systems science}, vol.~26, no.~10, pp.
  2015--2028, 1995.

\bibitem{hsieh2010optimality}
C.-S. Hsieh, ``On the optimality of two-stage kalman filtering for systems with
  unknown inputs,'' \emph{Asian Journal of Control}, vol.~12, no.~4, pp.
  510--523, 2010.

\bibitem{razi2018predictive}
A.~Razi, C.~Wang, F.~Almaraghi, Q.~Huang, Y.~Zhang, H.~Lu, and
  A.~Rovira-Sugranes, ``Predictive routing for wireless networks:
  Robotics-based test and evaluation platform,'' in \emph{Computing and
  Communication Workshop and Conference (CCWC), 2018 IEEE 8th Annual}.\hskip
  1em plus 0.5em minus 0.4em\relax IEEE, 2018, pp. 993--999.

\bibitem{drone-ADS-B}
\BIBentryALTinterwordspacing
``\textsc{ADS-B} transceivers, receivers and navigation systems for drones.''
  [Online]. Available:
  \url{http://www.unmannedsystemstechnology.com/company/uavionix-corporation/}
\BIBentrySTDinterwordspacing

\bibitem{dodge2009revealing}
S.~Dodge, R.~Weibel, and E.~Forootan, ``Revealing the physics of movement:
  Comparing the similarity of movement characteristics of different types of
  moving objects,'' \emph{Computers, Environment and Urban Systems}, vol.~33,
  no.~6, pp. 419--434, 2009.

\end{thebibliography}

\end{document}